\documentclass[%
 reprint,
superscriptaddress,
 amsmath,amssymb,
 aps,
]{revtex4-2}
\usepackage[utf8]{inputenc}

\usepackage{amsfonts}
\usepackage{upgreek}
\usepackage{graphicx}
\usepackage{hyperref}
\usepackage{dcolumn}
\usepackage{bm}

\usepackage{xcolor}
\usepackage[normalem]{ulem}

\newcommand{\um}{$\upmu$m }

\newcommand{\SiN}{SiN$_\text{x}$ }

\begin{document}

\title{
Coherent control of thermal transport with pillar-based phononic crystals}

\author{Tatu A. S. Korkiamäki}
 \affiliation{%
Nanoscience Center, Department of Physics, University of Jyvaskyla, FI-40014 Jyväskylä, Finland}%
\author{ Tuomas A. Puurtinen}%
\affiliation{%
Nanoscience Center, Department of Physics, University of Jyvaskyla, FI-40014 Jyväskylä, Finland}%
\author{Mikko Kivekäs}
\affiliation{%
Department of Physics, University of Jyvaskyla, FI-40014 Jyväskylä, Finland}%
\author{Teemu Loippo}
\affiliation{%
Nanoscience Center, Department of Physics, University of Jyvaskyla, FI-40014 Jyväskylä, Finland}%
\author{Adam Krysztofik}
 \affiliation{Faculty of Physics and Astronomy, Adam Mickiewicz University, 61-614 Poznan, Poland
  }
\author{Bartlomiej Graczykowski}
 \affiliation{Faculty of Physics and Astronomy, Adam Mickiewicz University, 61-614 Poznan, Poland
  }
\author{ Ilari J. Maasilta}
\email{maasilta@jyu.fi}
\affiliation{%
Nanoscience Center, Department of Physics, University of Jyvaskyla, FI-40014 Jyväskylä, Finland}%

\date{\today}

\begin{abstract}

Two-dimensional phononic crystals (PnCs) formed by a periodic array of holes in a suspended membrane have previously been used to coherently control thermal conductance at low temperatures by modifying the phonon dispersion, thereby altering the phonon group velocities and the density of states. Here, in contrast, we demonstrate that PnCs formed by a periodic array of Al pillars on an uncut \SiN membrane can also be used to achieve similar coherent control. We have measured and simulated the thermal conductance of four pillar-based PnCs with different lattice constants ranging from 0.3 to 5 $\upmu$m at sub-Kelvin temperatures, showing a strong up to an order of magnitude reduction in thermal conductance compared to an unaltered membrane. For the larger lattice constants $> 1 $ $\upmu$m, however, the experiments do not agree with the coherent theory simulations, which we interpret as a breakdown of coherence induced by increasingly effective diffusive scattering due to the roughness of the Al pillar surfaces. 
\end{abstract}

\maketitle

\section{Introduction}

A phononic crystal (PnC) is an artificial periodic structure in one, two or three dimensions that affects the propagation of phonons, the quanta of elastic waves\cite{Maldovan2013_review,Laude_2015_kirja}. Phononic crystals have been used to control elastic waves across a wide range of frequencies and wavelengths extending from seismic waves to hypersonic ($> 1$ GHz) range\cite{Vasileiadis2021}, where the operational principles rest on wave phenomena such as the formation of forbidden frequency bands (bandgaps) and the modification of dispersion relations. As phonons also carry heat, 
it is natural to ask whether the same wave phenomena can also be harnessed to control heat flow\cite{Maldovan_2015_review,Sledzinska2020,Nomura2022}. Such coherent effects in thermal conduction are more challenging to observe, as phonon coherence is more easily destroyed for high-frequency ($\sim 1$ THz) thermal phonons due to scattering from other phonons and structural imperfections\cite{Nomura2022}. Coherent thermal conduction has only been demonstrated conclusively for 1D superlattices with atomically smooth interfaces \cite{Luckyanova_2012,SuperlatticeCrossover}, and for 2D holey PnCs at low temperatures \cite{Zen_Comms_2014,Tian_maasilta_2019,Maire2017}, when the thermally dominant frequencies are much lower, down to the 1-10 GHz range in sub-Kelvin experiments\cite{Zen_Comms_2014}. Such lower frequency phonons have wavelengths larger than the typical experimental roughness of the surfaces, preventing diffusive scattering from such disorder that tends to destroy coherence at higher temperatures\cite{Ziman,Maznev_2015_correct_ziman}.     

Most previous experimental thermal conduction studies using PnCs have concentrated on geometries where a membrane or a beam is perforated by a periodic array of holes\cite{Nomura2022}. Such holey PnCs have been used to strongly reduce thermal conductance, over an order of magnitude in the low-temperature coherent limit \cite{Zen_Comms_2014, Tian_maasilta_2019}. However, such structures, having fragile suspended nanoscale features, are challenging to fabricate, in particular if one wants to use more advanced hole designs\cite{Florez2022}.   

Here, we focus on investigating coherent thermal conduction in a much less studied, but mechanically more robust alternative geometry for PnCs, based on the periodic arrangement of pillars on suspended membranes\cite{Pennec__2010_review,Anufriev__2018_pillar_review, Jin_2021} instead of holes. In holey PnC designs, the reduction of coherent thermal conduction is caused by wave interference due to scattering from the periodic holes, resulting in a strong modification of the phonon dispersion relations and, in turn, reduced phonon group velocities and densities of states, and in some cases formation of Bragg band gaps\cite{Zen_Comms_2014, Maldovan_2015_review,MaasiltaPuurtinen_2016_crystals,Tian_maasilta_2019,Anufriev_2016_coherent_diff_lattices}.
In contrast, in pillar-based PnCs a second additional coherent mechanism to modify the phonon dispersions is also effective: 
pillars can support local mechanical resonances, whose frequencies match those of the propagating modes of the underlying membrane at some values and can thus hybridize with them, causing hybridization band gaps at different frequency ranges from the Bragg gaps\cite{Pennec__2010_review,Laude_2015_kirja,Jin_2021,Khelif2010,Achaoui_2011_experimental_LR}. This also leads to additional flattening of the bands\cite{ Hsu_2011_local_reso,Khelif_2014-band-gap_physics,Hussein_2014,Xiong_2016}, 
as demonstrated experimentally by Brillouin light scattering experiments \cite{Sledzinska2020,Graczykowski_2015_dispersion,Yudistira_2016_pillar_BG_observation}. These strong modifications to the phonon band structure in pillar-based suspended PnCs are therefore also expected to lead to a significant reduction of the sub-Kelvin coherent thermal conduction\cite{Anufriev_2017_pilari}, the object of this study.  

  
Until now, nanoscale pillars-on-membrane PnCs have been used in experiments to guide GHz acoustic waves \cite{Poura2018}, improve the room temperature thermoelectric efficiency of Si\cite{Hussein_2023_pillar}, and reduce phonon thermal conductivity in ordered\cite{Anufriev_2017_Al_nanobeams} and disordered\cite{Anufriev_Nomura_2019_nanocones,Huang2020, Anufriev2023} pillar arrays at temperatures between 4 K and 300 K. The observed reduction in those thermal conductivity experiments has been modest, ranging from a few percent\cite{Anufriev2023} to $\sim 40\%$ \cite{Anufriev_Nomura_2019_nanocones}, depending weakly on the temperature. In almost all studies so far \cite{Anufriev_2017_Al_nanobeams,Anufriev_Nomura_2019_nanocones,Huang2020,Anufriev2023} the reduction has been attributed to incoherent scattering effects, as expected for higher temperature phonons, with the exception of one room temperature study, Ref. \cite{Hussein_2023_pillar}, where a role of the pillar resonance hybridization was suggested, however, without a quantitative agreement between the atomistic simulations and the experiment. Such coherence retaining atomistic molecular dynamics simulations of deep-nanoscale (period $< 10$ nm) pillars-on-membrane Si PnCs\cite{Wei_2015, Neogi2015,Hussein_2016_size-effects,Hussein_2018_2_order} predict much larger reduction factors, up to over two orders of magnitude\cite{Hussein_2018_2_order}, in disagreement with the experiments.

The only previous experiment on controlling thermal conductance using pillar structures at sub-Kelvin temperature range, to our knowledge, used microscale metal pillar arrays (periods 7 - 10 $\mu$m) on \SiN membranes and achieved a maximum reduction of 56\% at 0.1 K \cite{Zhang_2019_100mK}. The mechanism for the reduction in that study was not fully resolved but was attributed primarily to diffuse, incoherent scattering effects, likely arising from the electrons of the metal pillars. It is therefore fair to say that coherent, phonon dispersion-modifying effects of pillar-based PnCs have not been conclusively demonstrated in thermal conduction experiments before.

Here, we demonstrate experimentally that phonon thermal conduction can be strongly reduced coherently at sub-Kelvin temperatures, up to an order of magnitude, by 2D phononic crystals consisting of a periodic array of submicrometer to micrometer-scale pillars on a non-perforated suspended membrane. Numerical finite-element method (FEM) simulations of elastic waves in the measured geometries demonstrate a strong flattening of the dispersion relations of the thermally relevant phonon modes, which was confirmed by a Brillouin light scattering experiment. Such flattening was shown to correspond to a significant reduction in phonon group velocities, with a concomitant mild increase in the density of states, resulting in a net reduction in thermal conduction. The thermal conduction experiments on shorter-period structures up to one micron agreed well with the coherent simulations, whereas devices with larger dimensions deviated from the trends expected from coherent theory and more closely resembled the incoherent scattering picture.  

\section{Phononic crystal device design}

\begin{figure*}
    \centering
    \includegraphics[width=\textwidth]{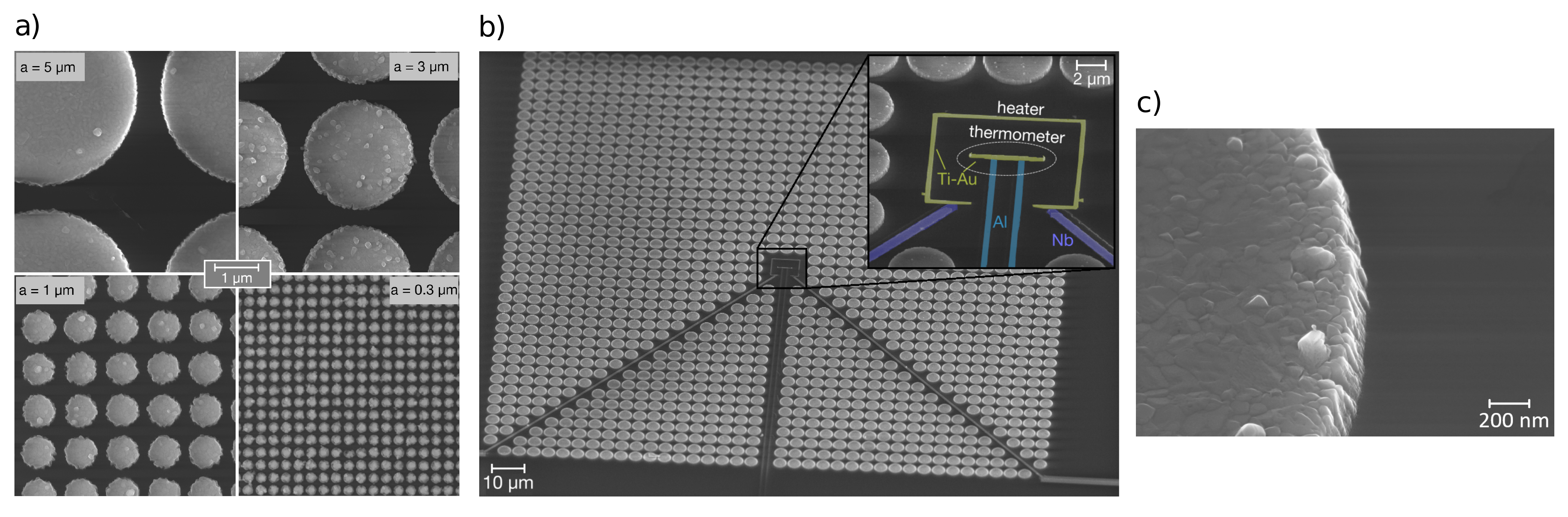}
    \caption{ \textbf{Pillar phononic crystal sample designs. (a)} Scanning electron microscope images of all measured PnC geometries with lattice constants 5, 3, 1, and 0.3 \um. The scale is the same for all four images. \textbf{(b)} A full view of a pillar PnC sample with a 5 \um lattice constant. The heater and the thermometer (identical for every sample) are highlighted in the zoom-in, with false colors to accentuate the different materials (yellow-Au, blue-Al, violet-Nb). \textbf{(c)} A high magnification image of the edge of an Al pillar, showing the surface and sidewall roughness, which is comparable in all samples, with rms values 8 nm and 70 nm, respectively.}
    \label{fig:SEm_PnC}
\end{figure*}

The experimental phononic crystal samples consist of 300 nm tall cylindrical polycrystalline Al pillars on a 320 nm thick amorphous \SiN membranes of size 300 $\times$ 300 $\upmu$m, suspended by etching through the underlying Si wafer using an anisotropic KOH backside etch. The pillar height was chosen to maximize the coherent effects on sub-Kelvin thermal conduction, based on previous simulations \cite{Anufriev_2017_pilari}. Four square lattice PnCs in total were measured, with lattice constants 0.3 $\upmu$m, 1 $\upmu$m, 3 \um and 5 $\upmu$m, all with an area filling factor of 0.65, corresponding to pillar diameters 0.27 $\upmu$m, 0.91 $\upmu$m, 2.73 \um and 4.55 $\upmu$m, respectively. A scanning electron microscope (SEM) image of  each PnC geometry is shown in Fig. \ref{fig:SEm_PnC} (a), with a full SEM image of the 5 \um sample shown in Fig. \ref{fig:SEm_PnC} (b). A metallic (Au) heater with superconducting Nb leads and an Al-AlO$_x$-Au superconductor-insulator-normal metal tunnel junction pair (SINIS) thermometer\cite{Giazotto2006,Koppinen2009,Zen_Comms_2014} are located at the center of the PnC structure (Fig. \ref{fig:SEm_PnC} (b)), used to perform the thermal conduction experiment. The pillars, heater and thermometer were fabricated on the suspended membrane using three rounds of electron-beam lithography and metal evaporation. For fabrication details, see Methods, Sect.\ref{fab}.

The superconducting Nb heater leads form direct low-resistance electrical contacts to Au, which fully confine the heating power to the normal metal part due to Andreev reflection\cite{Koppinen2009,Courtois}. Another notable difference in the heater-thermometer geometry from our previous PnC experiments\cite{Zen_Comms_2014,Maasilta_2016_bolometers,Tian_maasilta_2019} is that the thermometer is surrounded by the heater wire (Fig. \ref{fig:SEm_PnC} (b)), which in the ballistic phonon transport regime of the experiment leads to the equilibration of the thermometer and the heater temperatures (the view factor from the thermometer to the heater being close to one), without significant geometrical correction factors\cite{Tian_maasilta_2019} to influence the analysis (More details on the heater-thermometer geometry and materials in Methods sect. \ref{heat_thermo}). This has led to quantitative agreement between the coherent theory and the experiment for unmodified membranes, as will be shown below.           

The choice of the pillar material is based on the fact that Al is a superconductor at the sub-Kelvin temperatures $T < 0.6$ K of the experiment. In the superconducting state, electronic excitations are exponentially suppressed by the superconducting energy gap $\Delta$ at temperatures below the critical temperature $T_\textrm{C} = 1.2$ K \cite{Tinkham}, and small energy exchange inelastic diffusive phonon scattering from electronic excitations is therefore suppressed. In addition, the superconducting energy gap prevents phonons with energies below twice the gap energy $\hbar\omega < 2\Delta$ from breaking Cooper pairs, suppressing this second diffusive scattering channel for low-frequency phonons. By calculating the phonon number distribution for a 320 nm thick \SiN membrane at $T=0.6$ K, we estimate that practically all thermal phonons ($>$ 99 \%) have energies below $2\Delta = 0.34$ meV, the value for Al, demonstrating that the second inelastic phonon scattering channel is indeed negligible.     
For these reasons, the Al pillars act effectively as insulating mechanical extensions in the temperature range of this study, leaving surface and interface roughness (Fig. \ref{fig:SEm_PnC}(c)) as the main possible diffusive scattering mechanism for phonons.


\section{Results and discussion}

\subsection{Theoretical phonon band structure}

If phonons retain their coherence, Bragg scattering and mode hybridization strongly modify the phonon dispersion relations of the pillared PnCs. To determine the resulting new phonon eigenmodes for infinite periodic pillar square lattices, we have performed FEM simulations of the elasticity equations, assuming that both Al and \SiN have isotropic elastic properties (Methods, Sect. \ref{FEM}). The dimensions in the simulations followed the experimental samples and, more critically, the materials parameters used in the FEM simulations (density, Young's modulus, Poisson ratio) for both \SiN and Al were determined experimentally using Rutherford backscattering (Methods, Sect. \ref{Rutherford}) and Brillouin light scattering (BLS) (Methods, Sect. \ref{BLS}), to obtain more accurate band structure simulations.       

\begin{figure*}
    \centering
    \includegraphics[width=\textwidth]{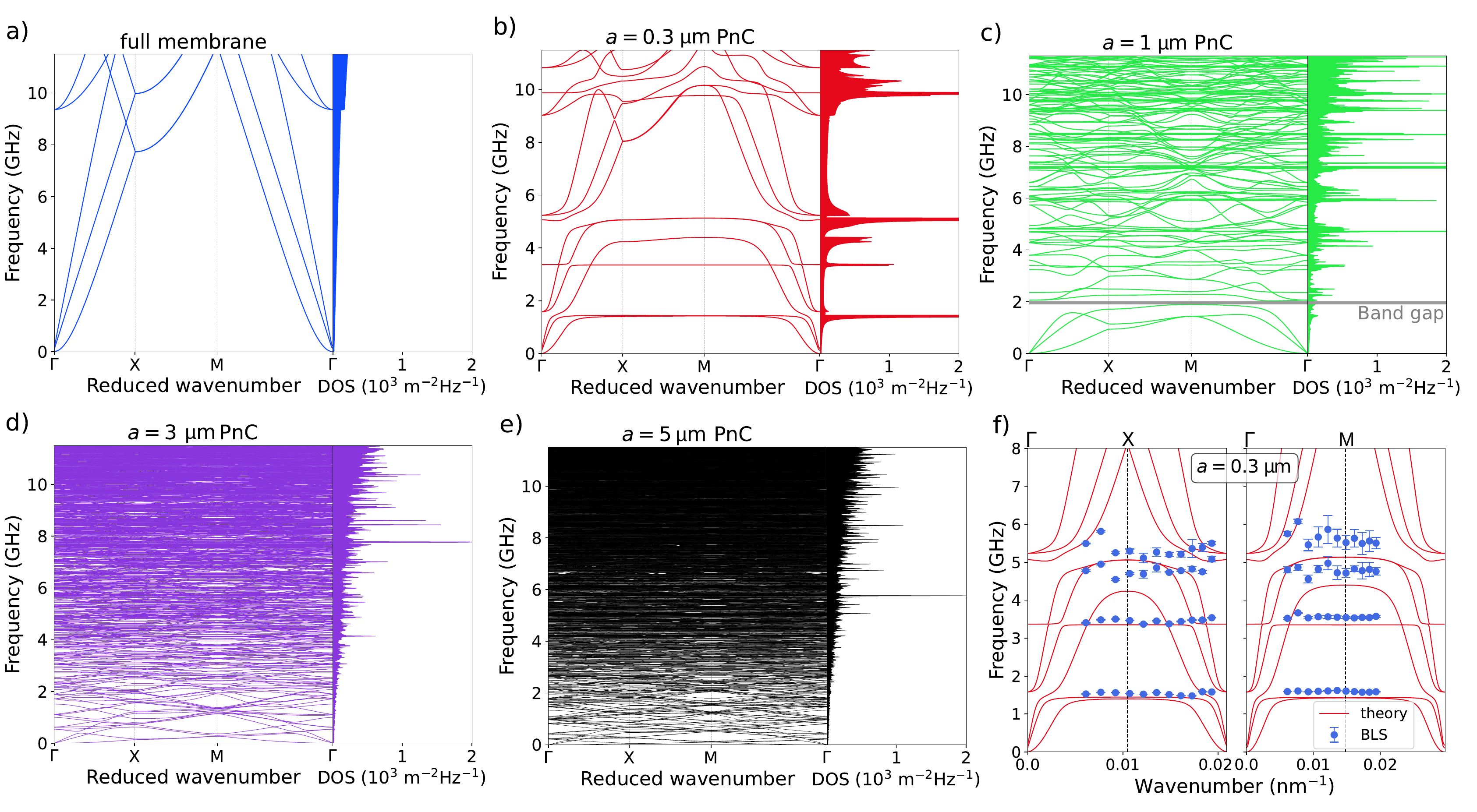}
    \caption{{\bf Band structure of the PnCs and comparison to BLS data.} The calculated phonon band structure $\nu(\bf{k})$ along the standard square lattice symmetry directions within the 1st BZ, for a \textbf{(a)} 320 nm thick \SiN membrane, and square lattice Al pillar PnCs (described in the text) with lattice constants  \textbf{(b)} 0.3 \um, \textbf{(c)} 1 $\upmu$m, \textbf{(d)} 3 $\upmu$m, and \textbf{(e)} 5 $\upmu$m. Each panel shows also the resulting density of states (DOS). \textbf{(f)} Comparison of  the simulated band structure to experimental data measured with Brillouin spectroscopy for the $a = 0.3$ \um PnC.}
    \label{fig:dispersion}
\end{figure*}

The calculated band structures (phonon frequencies $\nu({\bf k}) = \omega({\bf k})/2\pi$ as a function of the 2D wave vector $\bf{k}$) corresponding to all studied experimental pillar PnC designs with different periodicities (the lattice constant $a$ varying from 0.3 \um to 5 \um) are presented in Fig. \ref{fig:dispersion} and compared with the dispersions of an unmodified membrane, plotted in the same square lattice Brillouin zone symmetry directions as the PnCs. To allow for direct comparison, all band structures are plotted in the same frequency range up to 11.5 GHz. In addition to plots of the bands $\nu(\bf{k})$, we also plot the density of states (DOS) for each sample design.

Studying Fig. \ref{fig:dispersion}, we observe that the main effect of the pillars is that they generate flat bands (horizontal $\nu(\bf{k})$ portions), where each different flat band corresponds to a different vibrational resonant mode, localized mostly in the pillar portion of the structure rather than in the underlying membrane \cite{Graczykowski_2015_dispersion}. Such flattening of the bands has a two-fold effect on the properties that affect coherent thermal conduction\cite{Zen_Comms_2014}. On one hand, it leads to strong peaks in the DOS (see Fig. \ref{fig:dispersion} (b)), tending to increase thermal conductance, but on the other hand, it reduces the group velocities $\partial \omega/\partial {\bf k}$ around the resonant frequencies (all the way to zero in a fully flat band), which counteracts to reduce the conductance. It is therefore {\it a priori} not clear which effect in the end dominates without a more detailed discussion (Sect. \ref{G} below). In addition, a small band gap is observed for the 1 \um PnC (Fig. \ref{fig:dispersion}.(c)) around $\nu = 2$ GHz, but it is so narrow in frequency that its effect on thermal conductance is insignificant compared to the DOS and group velocity effects. 

Fig. \ref{fig:dispersion} also demonstrates that with increasing lattice constant, and therefore with increasing lateral pillar size (as we keep the pillar area fraction and volume fraction constant), many more flat band resonant modes appear within the frequency range plotted, creating an increasingly dense and on the average more flat band structure. 

\begin{figure*}
    \centering
    \includegraphics[width=\textwidth]{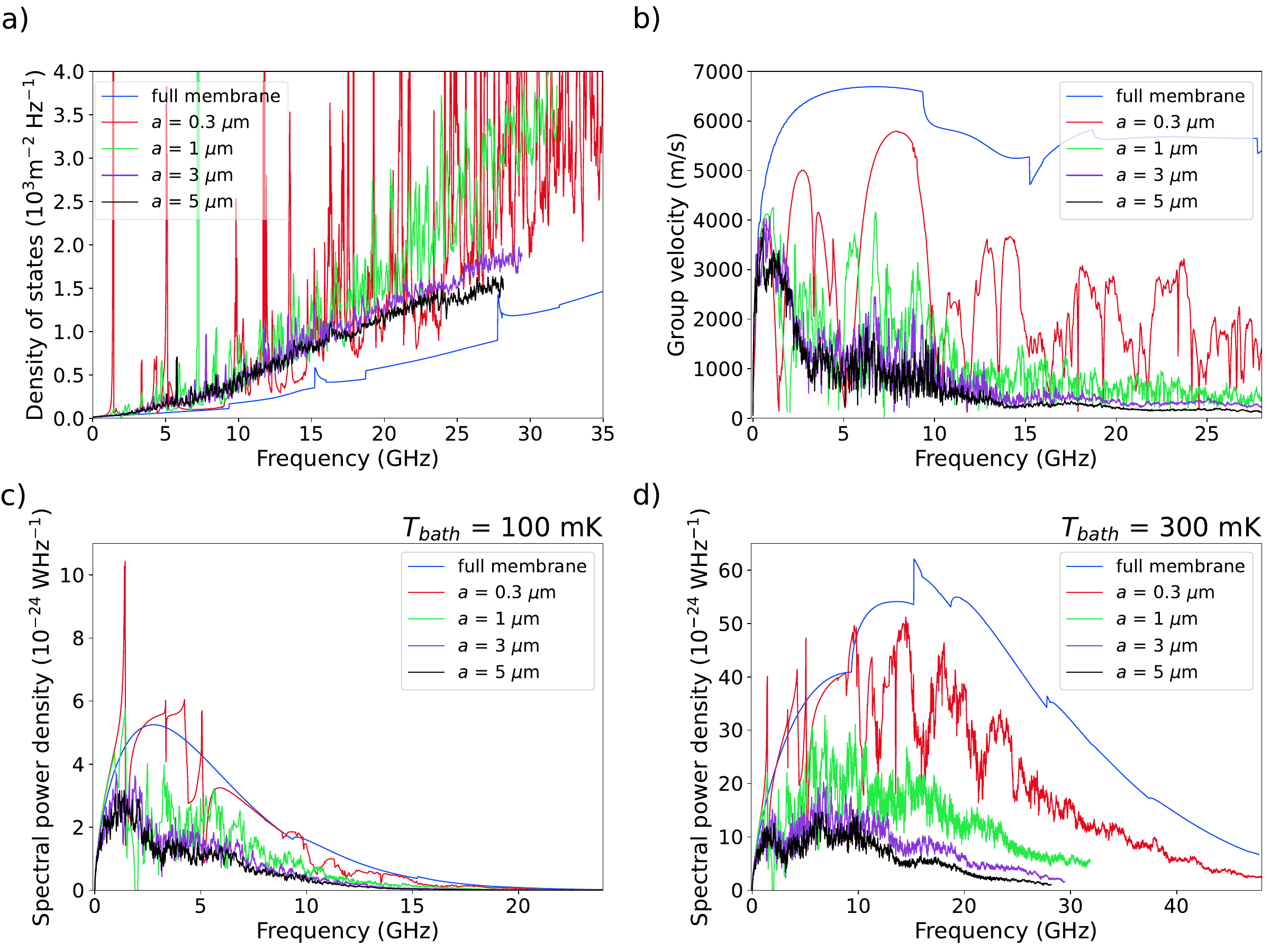}
    \caption{{\bf Spectral effects of the PnC band structures.} \textbf{a)} Phonon density of states and \textbf{b)} average spectral group velocity, calculated in the full 1st Brillouin zone for all PnCs and the unaltered membrane. The DOS data (apart from the full membrane) is smoothed for visual clarity. The calculated emitted spectral phonon spectral power density is presented  in \textbf{c)} at 100 mK and in \textbf{d)} at 300 mK for all the PnCs and the unaltered membrane.}
    \label{fig:DOS_yms}
\end{figure*}

To get a more complete picture on how the change in the lattice periodicity affects the DOS and the frequency-dependent average group velocity, we plot them for all PnC designs and for the unaltered membrane in Fig. \ref{fig:DOS_yms}, up to frequencies that contain essentially all the thermal phonons at 0.1 K \cite{Zen_Comms_2014}. The DOS (Fig. \ref{fig:DOS_yms} (a)) is clearly increased for all PnCs over the DOS of the unaltered membrane, but only by a factor of $\sim$ 2, on average. The PnC lattice constant $a$ does not seem to influence the overall scale of the DOS very much, the biggest difference is that for the smaller PnCs the fluctuations (peaks) in the DOS are much stronger, due to the sparser set of bands (Fig. \ref{fig:dispersion}). 

In contrast, the flattening of the bands leads to a strongly suppressed average group velocity, as seen in Fig. \ref{fig:DOS_yms}(b). This suppression keeps increasing with the periodicity, and is over an order of magnitude above 10 GHz for the larger period PnCs, reaching a factor as high as $\sim 300$ for the 5 \um period PnC at 28 GHz. For the smallest 0.3 \um period PnC, variations are again large as a function of frequency. It is also notable that the pillars have no effect on group velocity below $\sim 1$ GHz, where the long wavelength flexural modes dominate.  


\subsection{Experimental band structure determination with BLS}

To  experimentally demonstrate the existence of the modified modes in our pillar PnCs, and to investigate the accuracy of our FEM simulation model, we performed Brillouin light scattering (BLS) experiments on the smallest period $a = 0.3$ \um PnC design, to measure the lowest BLS-active phonon bands. The experimental details on the setup are described in more detail in Ref. \cite{Graczykowski_2015_dispersion}.   

In addition to the computational band structure, Fig. \ref{fig:dispersion} (f) includes the experimental dispersion relations measured with BLS, for the lowest BLS-active bands. The experimental BLS bands appear to match the simulated dispersion relations fairly well, especially with the two lowest bands, which are observed to be flat as the simulations predict. We also found that, for the smallest PnC structure, even the inclusion of a thin nm-scale native oxide layer on the surface of the Al pillars affected the simulated band structure, as such an AlOx cap stiffens the response. Thus, to achieve the best match between the data and the simulations, a  5 nm Al$_2$O$_3$ layer on all Al surfaces was included in the FEM model (See Sect. \ref{FEM} in Methods for details). The effect of such a thin oxide layer on the larger period structures was deemed insignificant because of the much lower volume fraction of the thin AlOx shell with the larger Al pillar volumes.

\subsection{Coherent thermal conduction modelling}
\label{G}

In the coherent limit, it is not possible to define a local thermal conductivity because of the lack of scattering mechanisms. Thermal conduction is then determined by the emitted power from a heater in analogy with the emission of thermal photonic radiation from a hot body. This phonon emission power can be calculated by integrating the product of the group velocity $\partial \omega_j(\bf{k})/\partial \textbf{k}$ and energy $\hbar\omega_j(\bf{k})$ for each emitted and thermally occupied phonon state ({\bf k},j), from each heater element. We stress here that the emitted power in that case depends on the geometry of the heater, as already demonstrated experimentally in Ref. \cite{Tian_maasilta_2019}. 

The heater model in our previous studies \cite{Zen_Comms_2014,MaasiltaPuurtinen_2016_crystals,Tian_maasilta_2019}, was based on the idea that the effective heater emission surface is perpendicular to the membrane plane, and is defined by the edges of the physical heater on the top membrane surface. In that model, it was also assumed that the phonon flux is emitted only outward, normal to the effective heater surface element, so that for the emission from narrow wires of length $L$, the emission power is proportional to $2L$. Here, we introduce a slightly modified model, where instead we assume that for a narrow wire, each small length element $dl$ of the heater wire can emit phonons in all directions with equal probability, without any possibility of self-reabsorption (shadowing) by the other elements of the heater wire.       
In that case, the emitted power can be obtained (for the derivation see Methods) from
\begin{equation}
        P(T) = \frac{L}{(2\pi)^2}\sum_j \int_K d\textbf{k} \hbar \omega_j(\textbf{k})n_\textrm{B}(\omega,T)\left|\frac{\partial \omega_j(\textbf{k})}{\partial\textbf{k}}\right|,
    \label{eq:emitted_power_uus}
\end{equation}
 where $\omega_j(\textbf{k})$ are the phonon dispersion relations for each branch $j$, $K$ is the first Brillouin zone of the PnC and $n_B(\omega,T)$ is the Bose-Einstein distribution at a heater temperature $T$. 

For the unmodified isotropic membrane, in Eq. \ref{eq:emitted_power_uus}, the k-space integral can be written in polar coordinates over the entire 2D k-space, allowing a direct comparison of Eq. \ref{eq:emitted_power_uus} with the older model in Refs.\cite{MaasiltaPuurtinen_2016_crystals,Tian_maasilta_2019} after performing the angular integrals. The difference between the two models is then just a constant, where $P_{\textrm{new}}=\frac{\pi}{2} P_{\textrm{old}}$, for all materials and thicknesses of the membrane. This difference is quite important when we compare the coherent modelling to the experimental results below.
We also note that for the unmodified \SiN membranes, FEM simulations need not be used to calculate the dispersion relations: instead, we use the Rayleigh-Lamb theory for elastic plates\cite{Graff_Elastic_waves,Kuhn_2006} to simplify the numerics significantly.   

\begin{figure*}
    \includegraphics[width=\textwidth]{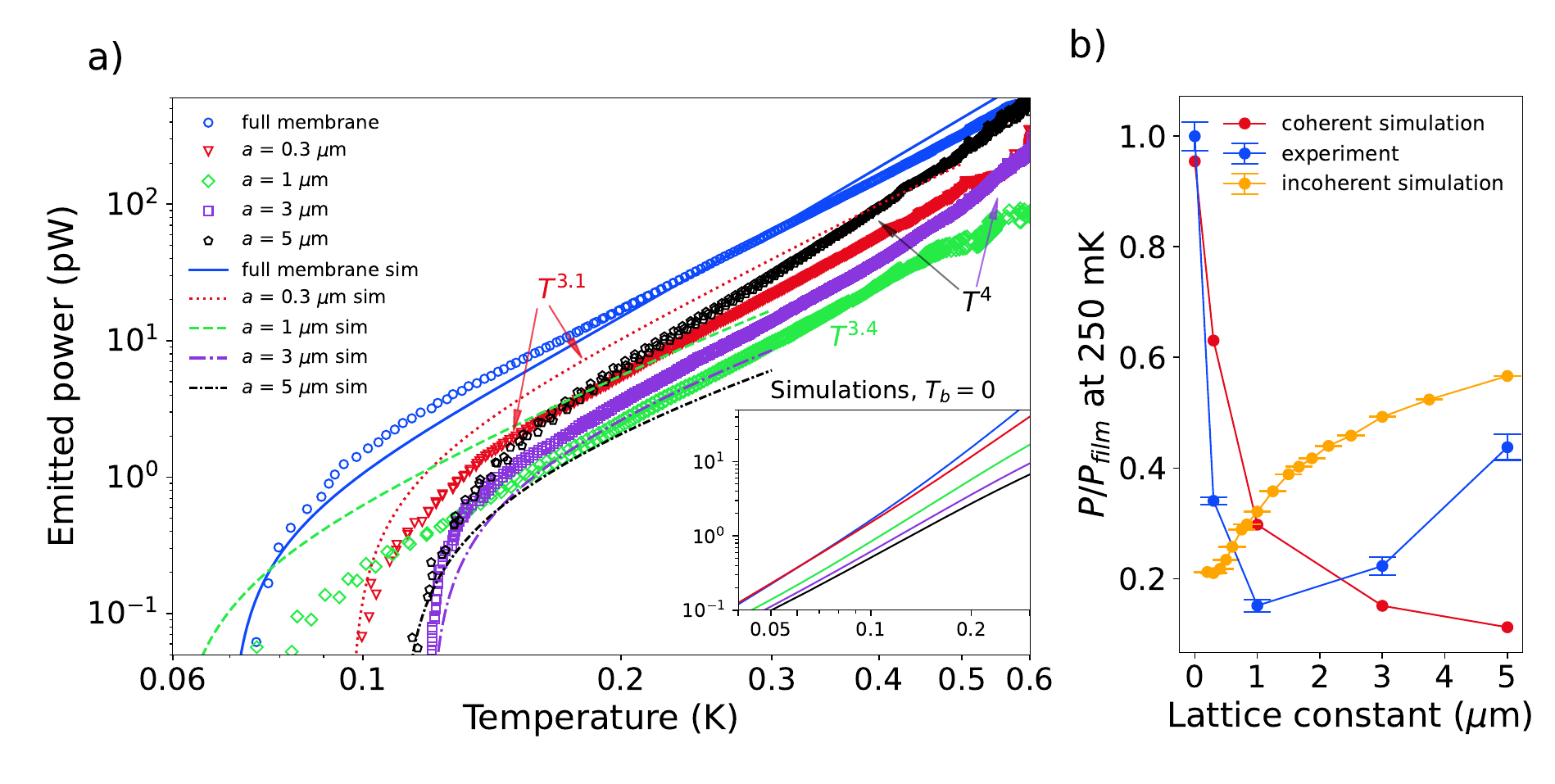}
  \caption{\textbf{Experimental emitted phonon power and comparison to theory. a)} Experimental and coherent theory simulation results for an unaltered 320 nm \SiN membrane as well as Al pillar PnCs with lattice constants 0.3, 1, 3, and 5 $\upmu$m. The inset presents the coherent simulations again for a better visual clarity with the same color coding without taking into account the bath temperature of the experiment. \textbf{b)} Comparison of emitted power at 250 mK relative to the unaltered membrane for the experimental data, the coherent theory, and the Monte Carlo simulations. The lines are guides to the eye.}
    \label{fig:emitted_power} 
\end{figure*}

Before looking at the experimental data, we can first look at the theoretical emitted spectral power density $P(\nu)$ ($P = \int P(\nu)d\nu$) for all experimental sample PnC designs and the unmodified membrane, plotted in Fig. \ref{fig:DOS_yms}(c) and (d) for two temperatures $T = 0.1$ K and $T = 0.3$ K, respectively. The emitted phonon power is clearly lower for the PnCs than for the unmodified membrane, practically across the entire spectrum, the difference increasing with higher temperatures. This demonstrates that the effect of the decrease in the group velocities is in general stronger than the  effect of the increased DOS, leading to an overall decrease in power density. However, the smallest period 0.3 \um PnC at the lower 100 mK temperature is the only example where the phonon power spectrum roughly maintains the shape and magnitude of the full membrane spectrum. With the sizeable peak at around 1.5 GHz corresponding to the states near the lowest flat band in the band structure in Fig. \ref{fig:dispersion} (b), we can even expect an enhancement in thermal conductance at temperatures below 100 mK. At higher temperatures the 0.3 \um PnC behaves more similarly to the other PnCs, a clear decrease already being visible at 300 mK in Fig. \ref{fig:DOS_yms} (d).


\subsection{Experimental thermal conduction results and comparison to theory}


The emitted phonon power in the experiment and from the coherent theory simulations for the Al pillar-based PnCs, as well as for the unmodified SiN membrane are presented in Fig. \ref{fig:emitted_power} (a), as a function of the temperature of the heater (see Sect. \ref{meas_details} for the measurement details). At low power, the temperature naturally saturates to the bath temperature $T_{\textrm{bath}}$ of the particular experimental run, which varied between 60 mK - 120 mK.  In the theoretical curves, this is taken into account by including a back emission power term such that the net emitted power is $P_{\textrm{net}}(T,T_{\textrm{bath}}) = P(T)-P(T_{\textrm{bath}})$, where both terms are calculated with Eq. \ref{eq:emitted_power_uus}, without {\em any} adjustable fitting parameters. To make the inherent theoretical power laws apparent also at the low temperature limit, the inset of Fig. \ref{fig:emitted_power} (a) shows the theoretical $P(T)$ curves with the effect of the bath temperature removed, i.e. with $T_\textrm{bath} = 0$.   

The first thing to notice is that for the unmodified, full membrane, the coherent simulation and the experiment agree quite well, even quantitatively. This gives us evidence that the new heater design and modelling behind Eq. \ref{eq:emitted_power_uus} are improvements. Turning next to the PnC data, the experiments with the 0.3 \um and 1 \um period PnCs seem to qualitatively agree with the coherent theory, which predicts a monotonic decrease in conductance as a function of the lattice constant (see Fig. \ref{fig:emitted_power}(b)). In particular, the experimental temperature dependence of the 0.3 \um crystal $P \sim T^{3.1}$ is the same as for the coherent simulation, although the absolute value is a factor of $\sim 2$ below the coherent simulation. 
A similar difference in absolute values exists for the 1 \um PnC, for which the experimental power law is now $T^{3.4}$, 
somewhat higher than the corresponding one in the simulation, $T^{2.8}$. More importantly, the emitted power of the 1 \um PnC is close to an order of magnitude lower than the emitted power of the unaltered membrane, making this the highest experimentally observed thermal conductance reduction achieved with pillar-based PnCs to date, to our knowledge. 

To quantify this reduction in terms of thermal conductance $G = P/(T - T_\textrm{bath})$ that can be compared to previous studies, we calculate the value of specific thermal conductance $G/L$ (independent of the heater length $L$) for the 1 \um PnC, giving $G(0.2 \, \textrm{K}) =  0.52$ pW/(K$\upmu$m) at 0.2 K. This value is only about a factor of $\sim 2$ higher than the previous record achieved with holey 2D PnCs\cite{Tian_maasilta_2019}, demonstrating the high effectiveness of pillar-based PnCs for the reduction of thermal conductance.   

On the other hand, for the larger lattice constants of 3 \um and 5 $\upmu$m, the experimental conductance is higher than the prediction of the coherent theory. It fails to predict even the trend seen in the experimental results, where the thermal conductance instead starts to rise with an increasing lattice constant (Fig. \ref{fig:emitted_power} (b)). We have previously observed similar behaviour in hole-based PnCs, where the turnaround of the decreasing trend occurred with a slightly larger $a$ \cite{Tian_maasilta_2019}. The 3 \um PnC seems to follow a $T^{3.1}$ power law up to around 300 mK, after which the exponent changes to $T^4$. For the largest period 5 \um PnC, the exponent of the power law remains at 4 for the entire measurement range. This $P \sim T^4$ dependence corresponds to regular 3D bulk phonon modes being the heat carriers, as opposed to coherently modified PnC modes. We also note that this power law has been shown to be valid not only for fully ballistic bulk phonons \cite{Kuhn_2007_maximising}, but also in the Casimir limit of diffusive surface scattering for heat flow in the membrane geometry\cite{TuomasAIP}. Thus, the data for the longer period PnCs strongly support the idea of the breakdown of coherence due to diffusive scattering, most likely at the rough surfaces.

To study the incoherent limit further, we have performed Monte Carlo simulations using bulk 3D modes for varying PnC lattice constants corresponding to the experimental geometry (for details see Methods Sect. \ref{MC}), which predict a monotonic increase in thermal conductance as a function of the lattice constant, as presented in Fig. \ref{fig:emitted_power} b). This strengthens the conclusion for the breakdown of coherence as a function of the PnC period, which starts to take place at $a \approx 1 \, \upmu$m. This breakdown is most likely the result of diffusive scattering from the rough Al pillar surfaces and grain boundaries within the pillar. 
Finally, the dominant phonon wavelength for Lamb modes in a 300 nm thick \SiN membrane is $> 1$ \um for $T < 200$ mK \cite{Tian_maasilta_2019}. Therefore,  the smaller 0.3 \um and 1 \um period pillars can behave more like point-like scatterers, which most likely lowers the likelihood of diffusive scattering that destroys coherence.  


\section{Conclusions}

We have measured the thermal conductance of four pillar-based phononic crystals at sub-Kelvin temperatures, demonstrating a reduction by up to an order of magnitude. We attribute this reduction mostly to a coherent modification of the phonon band structure, which significantly lowers the phonon group velocities. Brillouin light scattering experiments also provided direct evidence for the coherently modified dispersions of the smallest-period device, and thermal conduction simulations corroborated the coherent picture by qualitatively agreeing with the thermal conduction experiments.  

For the larger-period crystals, the agreement with coherent theory appeared to progressively vanish, indicating a transition to an incoherent transport regime. Most likely, this breakdown of coherence is due to diffusive surface scattering from the rough pillar walls, which is more effective for larger pillar sizes.   

To reduce the effect of roughness, we are developing fabrication methods for pillars with smoother sidewalls. Another future possibility to achieve an even stronger coherent thermal conduction reduction is by reducing the underlying membrane thickness \cite{Anufriev_2017_pilari}. In the future, pillar-based PnCs can provide an alternative to their hole-based counterparts in low-temperature applications, for example in ultrasensitive radiation detection and quantum technology, especially in conditions requiring a much more mechanically robust structure.

\section{Methods}

\subsection{Sample fabrication}
\label{fab}

All samples were fabricated on commercial $\langle 100\rangle$ Si chips coated on both sides with 320 nm thick low-stress amorphous \SiN films grown by low-pressure chemical vapour deposition (LPCVD), purchased from the Microfabrication Laboratory at UC Berkeley. Suspended \SiN membranes were created by patterning the backside of the chip to a large square via positive e-beam lithography with a Raith eLine SEM using a PMMA resist. The exposed \SiN was then removed by a reactive ion etching (RIE) process with a CHF$_3$/O$_2$ mixture (10:1) as a process gas, 100 W RF power and 55.0 mTorr pressure at  $30 \,^\circ$C (Oxford Instruments Plasmalab 80). A standard anisotropic crystallographic Si etch process with KOH removed the substrate material, creating suspended \SiN films of around 300 $\times$ 300 $\upmu$m area on the top surface. A dilute solution of HNO$_3$ ($\sim 0.1 \%$) was used to clean the chip of any residual materials left from the etching solution.

Next, the PnC pillars were fabricated onto the membrane. E-beam lithography using a roughly 1000 nm-thick layer of PMMA 950 resist was employed to pattern circular holes with the desired pillar diameter in the resist. The resist was spin coated in three layers, with a 5 nm layer of Al evaporated between the second and third layer to prevent charging effects, which can be quite severe in the suspended insulating membranes. After development, any remaining PMMA residues on the exposed \SiN surfaces were removed with a soft O$_2$ plasma cleaning (60 W, 40 mTorr, 30 s) in the RIE. The 300 nm tall Al pillars were deposited in a e-beam UHV evaporator (base pressure $\sim 10^{-8}$ mbar) with a 0.1 nm/s evaporation rate, from an angle normal to the substrate. Lift-off was performed in a warm acetone bath.

The heater and thermometer were deposited in the middle of the PnC structure, with leads extending outside of the membrane area (Fig. \ref{fig:SEm_PnC} (a)) to bonding pads, in two separate fabrication rounds, using the same UHV e-beam evaporator as before. E-beam lithography was again used, this time with a multilayer resist: A copolymer P((8.5)MAA)MMA in 9 \% ethyl lactate at the bottom (400 nm) to create an undercut for shadow evaporation, and two layers of PMMA 950 on the top (350 nm each) with a 5 nm layer of Al in between. Here, the Al not only prevents charging but also prevents the formation of cracks in the resist around the sharp edges of the structure. The Al layer between the PMMA layers will also preserve the good contrast required for a three-point alignment procedure, performed on previously fabricated markers to create the pattern at the correct spot in the middle of the pillars. If the conductive Al were on top, the markers would not be visible. RIE was again used for plasma cleaning residues after the development. The 500 nm wide superconducting Nb heater leads were fabricated first, for which 5 nm of Ti and 30 nm of Nb were evaporated along the first lead direction from $60 ^\circ$ angle with respect to the surface normal, with evaporation rates of 0.1 nm/s and 0.2 nm/s, respectively. This direction of evaporation was used to create the contact between the Nb lead and the Au normal metal, and to prevent any extra Nb features from appearing because of the normal metal pattern in the resist. The sample stage was then rotated by $90 ^\circ$ around the surface normal, and the process was repeated for the second Nb lead direction, resulting in the two diagonal Nb leads visible in Fig. \ref{fig:SEm_PnC} (a). After the leads, the rectangular (dimensions 10 $\upmu$m $\times$ 8 $\upmu$m, with a total heater length of 32 $\upmu$m)  Au normal metal wire (width 300 nm) portion of the heater that emits the thermal phonons was deposited normal to the surface without breaking the vacuum, using 5 nm of Ti and 30 nm of Au (rate 0.1 nm/s).  

The lithography steps for the SINIS tunnel junction thermometer fabrication were the same as for the heater, the only difference being the absence of the copolymer layer. The evaporation began with the deposition of the $\sim$300 nm wide superconducting leads, by evaporating 25 nm of Al (0.1 nm/s) from 60$ ^\circ$ angle along the direction of the leads, thus forming Al wires only in the desired places without a feature from the normal metal pattern in the resist. The Al was then oxidised in pure oxygen at 200 mbar for 5 min in the load lock of the evaporator. After the oxidation, the stage was rotated 90$  ^\circ$ around the surface normal, and inserted back into the evaporation chamber for the steps to deposit the $\sim$300 nm wide Ti-Au normal metal wire. First Ti and then Au were evaporated from both sides along the normal metal pattern (perpendicular to the Al leads, see Fig. \ref{fig:SEm_PnC} (a)) first from $ 55  ^\circ$, then $ 50  ^\circ$ and finally from $ 45^\circ$ angles, at a rate of 0.1 nm/s. Angled evaporation again guarantees that no extra Au-Ti features appear on the device, and the use of multiple angles ensured even coverage of the oxidized Al. 
The obtained bottom Ti adhesion layer thickness was in the end $\sim$ 5 nm, with the Au layer being $\sim$35 nm thick. For the junctions fabricated this way, typical room temperature resistances are in the range 4-8 k$\Omega$.

\subsection{Heater-thermometer geometry and materials}
\label{heat_thermo}

Our previous PnC thermal conductance measurements have been conducted with SINIS-junctions as both the heater and the thermometer, fabricated "face-to-face", with the two linear normal metal sections facing each other \cite{Zen_Comms_2014,Maasilta_2016_bolometers,Tian_maasilta_2019}. However, data-analysis for this geometry is complicated by the fact that not all heat emitted from the heater into the half-space towards the thermometer is intercepted by it, causing the heater and thermometer temperatures to slightly differ, which leads to a geometrical correction factor for the power versus temperature curves\cite{Tian_maasilta_2019}. In addition, as the SINIS-heater resistance is dominated by the tunneling resistance (as opposed to just the resistance of the normal metal island), practically all of the heat is generated by the tunneling processes, half of which is dissipated in the superconducting leads. 

To account for these non-idealities, a new measurement geometry was developed, where the heater is a normal metal wire with direct, electrically low-resistance contacts to superconducting leads (SNS heater), with the normal metal portion encircling the SINIS-thermometer (Fig. \ref{fig:SEm_PnC} (a), inset). Without tunnel junctions in the heater part, phonon radiative thermal emission is now uniformly emitted only from the normal metal portion of the heater, as Andreev reflection processes at the normal-superconductor contacts transform the dissipative current in the normal metal into a non-dissipative supercurrent in the superconductor. In addition, as the thermometer is surrounded by the heater almost completely, it "sees" almost all the radiative power emitted into the inner portion of the rectangular region defined by the normal metal heater (the view factor from the thermometer to the heater being close to one). In such a geometry,  the steady state conditions of radiative power balance cause the thermometer temperature to equilibrate to the heater temperature without significant geometric correction factors.       

The normal metal part of the heater consists of a Ti-Au bilayer, with the lower $\sim 5$ nm thick layer of Ti acting as an adhesive between the \SiN membrane and the $\sim 30$ nm thick gold wire. This thin layer of Ti was not seen to exhibit superconductivity within the temperature range of the experiment (down to $\sim$ 50 mK) due to the inverse proximity effect of the normal Au metal layer in good electrical contact with the Ti layer. The superconducting leads in the heater are 30 nm-thick niobium, possessing a high critical current and critical temperature ($\sim 6$ K) and a large superconducting gap. In the sub-Kelvin temperature range of the experiment, such leads act as excellent electronic thermal insulators\cite{Koppinen2009,Hoffmann2002}, that is, hot electrons from the normal metal region cannot diffuse into the leads. This combined with the low value of contact resistance of the contacts between Ti-Au and Nb (absence of a tunneling barrier) allows for a large biasing range for the heating experiment.  

 As our experiment does not extend above 1K, the best choice for a superconducting material for the SINIS thermometer is Al, as high-quality NIS junctions can be fabricated consistently with it\cite{Giazotto2006,Koppinen2009,Zen_Comms_2014}. For the normal metal island, we have also used the same Ti-Au bilayer as for the heater, as it has been found to be superior to copper  in these applications \cite{Ilmo_2018_Ti-Au,Tian_maasilta_2019}. The advantages of Ti-Au compared to Cu as a normal metal include greater durability, both in terms of chemical attacks during the fabrication and a smaller risk of breakdown due to static charges, the latter of which is caused by a lower specific tunnelling resistance. The reduced resistance also reduces self heating in the thermometer. On the other hand, a Ti-Au junction may exhibit noticeable self-cooling \cite{Giazotto2006} due to the lower tunnelling resistance, which can have an effect on it's accuracy for reflecting the true phonon temperature at the lower temperature regime of the experiment (see section \ref{meas_details}.)

\subsection{Measurement details}
\label{meas_details}

Low temperature thermal conduction measurements were performed in a $^3$He-$^4$He-dilution refrigerator with a base temperature $T_\textrm{bath}$ $\sim 50$ mK, with multiple stages of electrical filtering present on the measurement lines at different temperatures. 
At the sample stage at base temperature, there are RC low-pass filters in addition to a home made continuous Eccosorb filter to absorb microwave noise. At 4 K, the lines have commercial low-pass pi-filters. Additionally, the voltage and current preamplifiers used (Ithaco 1201 and Ithaco 1211, respectively) also possess their own output filters, set for low-pass/DC for the measurements.

Prior to the actual thermal conductance measurement, the SINIS-junction thermometer was characterised with a set of current-voltage (IV) measurements, on the basis of which a suitable bias current for thermometry was chosen. With a constant bias current, the voltage of the SINIS-junction is a sensitive measure of the temperature of the normal metal island below the $T_\textrm{C}$ of the superconductor\cite{Giazotto2006,Koppinen2009,Zen_Comms_2014}. A calibration of the SINIS thermometer was performed by simultaneously measuring the voltage across the SINIS  with the resistance of a calibrated RuOx thermometer (located on the sample stage) during a slow cooldown of the refrigerator from $\sim 1$ K to the base temperature. Self-heating of the junction was minimized by using a low bias current well within the superconducting gap.     


In the following thermal conduction experiment, the emitted power from the heater and the resulting temperature increase were measured as a function of the bias voltage across the heater. The voltage of the SINIS thermometer and the voltage and the current of the heater were recorded, with the temperature obtained using the calibration data and the emitted Joule heating power as $P = IV$. As the leads of the normal metal heater were superconducting niobium, only the normal metal region of the heater was subject to heating.

\begin{figure}
    \centering
    \includegraphics[width=0.9\linewidth]{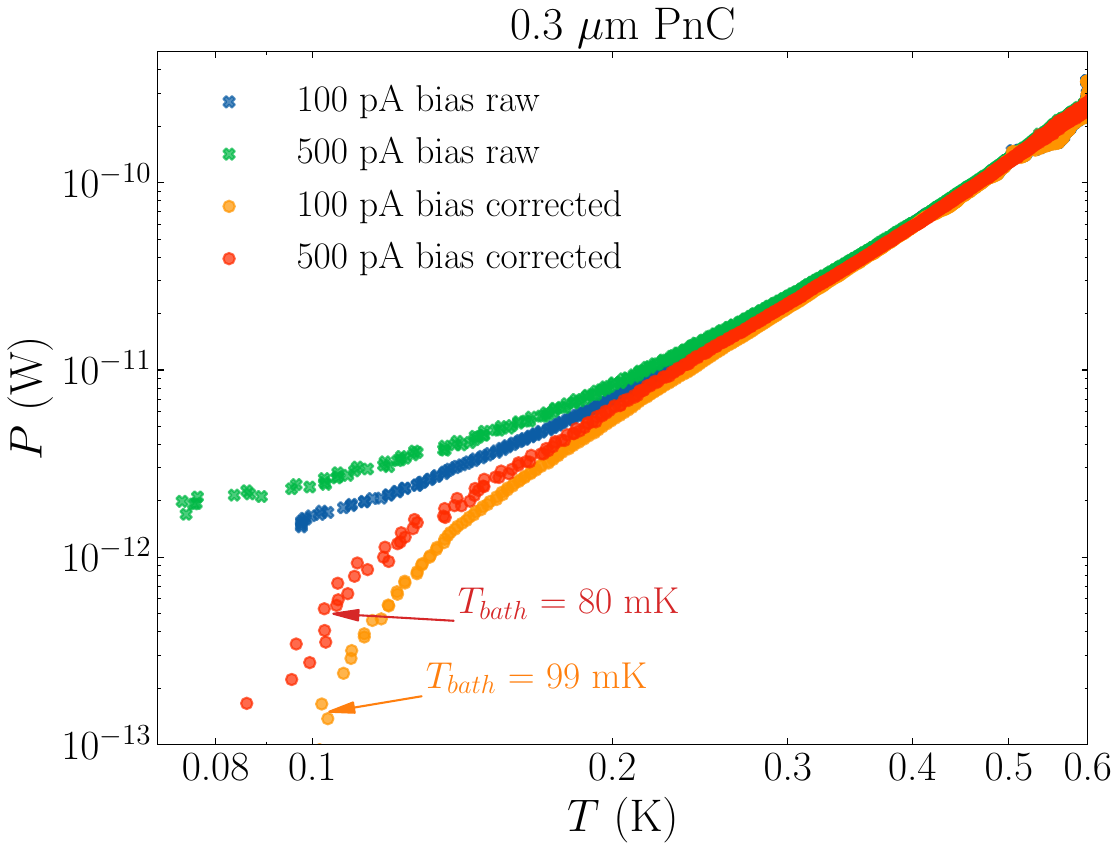}
    \caption{{\bf The effect of self-cooling on the thermometry.} The 0.3 \um phononic crystal emitted power data with and without correcting for SINIS-cooling. With a larger bias, the cooling power is greater.}
    \label{fig:300nm_raw}
\end{figure}

For two samples with the lowest tunneling resistance a self-cooling effect of the SINIS junction thermometer\cite{Giazotto2006,Chaudhuri_maasilta_2012} was visible in the emitted power data. The two measurements where this was observed were the 0.3 \um PnC and the full membrane measurements. The tunneling resistances of the thermometers for these samples were 3.9 k$\Omega$ and 3.2 k$\Omega$, respectively. The resistances for the rest of the junctions were between 4.4 and 5.8 k$\Omega$, and for these no clear cooling effect was observed. Fig \ref{fig:300nm_raw} presents the effect: The low temperature tail turns upwards, instead of going to zero at the bath temperature. The effect is stronger with a larger current bias, as the current also increases the voltage over the junction pair. The effect can be corrected for with a constant term in the order of $10^{-12}$ W obtained from a power law fit. In reality, the temperature dependence of the SINIS cooling is not a constant; it is dependent on the temperature and the voltage over the junction, which is also dependent on temperature when biased for thermometry. Regardless, due to the power law nature of the $P(T)$ curve, the small constant correction becomes rapidly negligible at higher temperatures.


\subsection{FEM simulations}
\label{FEM}

The phonon band structures (dispersion relations $\omega(\bf{k})$) of periodic square lattice pillar-PnC structures, corresponding to all four experimental designs, were calculated by solving the 3D elasticity equations for isotropic materials \cite{Zen_Comms_2014,Graff_Elastic_waves} using the finite element method (FEM). 
Periodic Bloch-wave boundary conditions were used with 2D wave-vectors \textbf{\textit{k}} in the x-y plane, with typically $\sim$ 700 - 3200 k-points in the irreducible octant of the first Brillouin zone. Depending on the value of the PnC period, $\sim 1 700 - 21 000$ eigenvalues were computed, extending the frequency range to at least $\sim 30$ GHz for all periodicities. Results for the SiN membrane dispersion relations without pillars were also calculated using the Rayleigh-Lamb theory \cite{Graff_Elastic_waves}. 

In the simulations, we used the experimentally determined material parameters (Sects. \ref{Rutherford} and \ref{BLS} below) for both the LPCVD-grown SiN (Young's modulus $E = 253.5$ GPa, Poisson ratio $\nu = 0.24$, and density $\rho = 2870$ kg/m$^3$) and the evaporated polycrystalline Al ($E = 70$ GPa, $\nu = 0.34$, and $\rho = 2700$ kg/m$^3$). In addition, from the simulations we observed that for the smallest period PnC with $a = 0.3$ $\upmu$m, the addition of just a few nm thick Al oxide layer on the surface of the Al pillars affects the elastic response and therefore the band structure details. Thus, to fit the experimental BLS spectra, a 5 nm Al$_2$O$_3$ layer on the Al pillar surfaces ($E = 340$ GPa, $\nu = 0.22$, and $\rho = 3970$ kg/m$^3$, literature values for Al$_2$O$_3$ \cite{CRC}) was used in the simulations for the smallest dimension PnC. For the larger structures, the effect of an oxide layer was deemed insignificant, as the volume fraction of the oxide for them is small $< 4$ \%, much smaller than the $\sim 9 $ \% for the smallest period PnC. Moreover, even for the smallest period PnC (0.3 µm), the effect of adding the oxide only resulted in a difference of $< 2$ \% in thermal conductance.


\subsection{Incoherent Monte Carlo simulations}
\label{MC}

The FEM simulations assume all effects to be coherent and all scattering to be specular. Therefore, it does not take into account any diffusive scattering, particularly in the relatively rough surfaces of the aluminium (Fig. \ref{fig:SEm_PnC} b). To account for this, we performed Monte Carlo ray tracing simulations \cite{PhononRT} to simulate the thermal conductance in the incoherent regime. As in a previous study \cite{Tian_maasilta_2019}, we consider phonons as particles, and simulate the phonon transmission probability through the pillar PnC structure with the same membrane thickness, filling factor and pillar heights as the experimentally measured samples, with lattice constants $a$ varying in the experimental range between 200 nm and 5 \um, while keeping the overall dimensions of the structure (15 \um $\times$ 15 $\upmu$m) constant. The conductance and emitted power is obtained from the transmission probability using the Landauer-B\"uttiker formalism \cite{Hori_2015_Landauer,Jeong_2015_laundauer,Tian_maasilta_2019}. 

As an improvement to previous simulations, we also account for phonon mode conversions \cite{Auld} affecting the reflections and refractions at the surfaces and interfaces. The probability for diffusive scattering was obtained using a specularity parameter $p$ for each type of surface/interface. For the surface specularities, we assume the \SiN to be fully specular ($p = 1$). For the Al pillars, the top RMS surface roughness was measured with atomic force microscopy to be 8 nm. Using the Ziman-Maznev formula for the specularity parameter \cite{Ziman,Maznev_2015_correct_ziman}
\begin{equation}
        p(\lambda) = \exp\left( -\frac{16 \pi^2\eta^2}{\lambda^2}\right),
        \label{eq:ziman}
\end{equation}
where $\eta$ is the RMS roughness, we estimated the specularity for the dominant phonon wavelength in Al at 250 mK, corresponding to $p = 0.96$. For the considerably rougher pillar sidewalls, $p = 0.22$ $(\eta = 70 $ nm) was estimated based on SEM images, whereas for the bottom interface between the pillars and SiN$_\text{x}$, $p = 0.99$ was used.

\begin{figure}
    \centering
    \includegraphics[width=\linewidth]{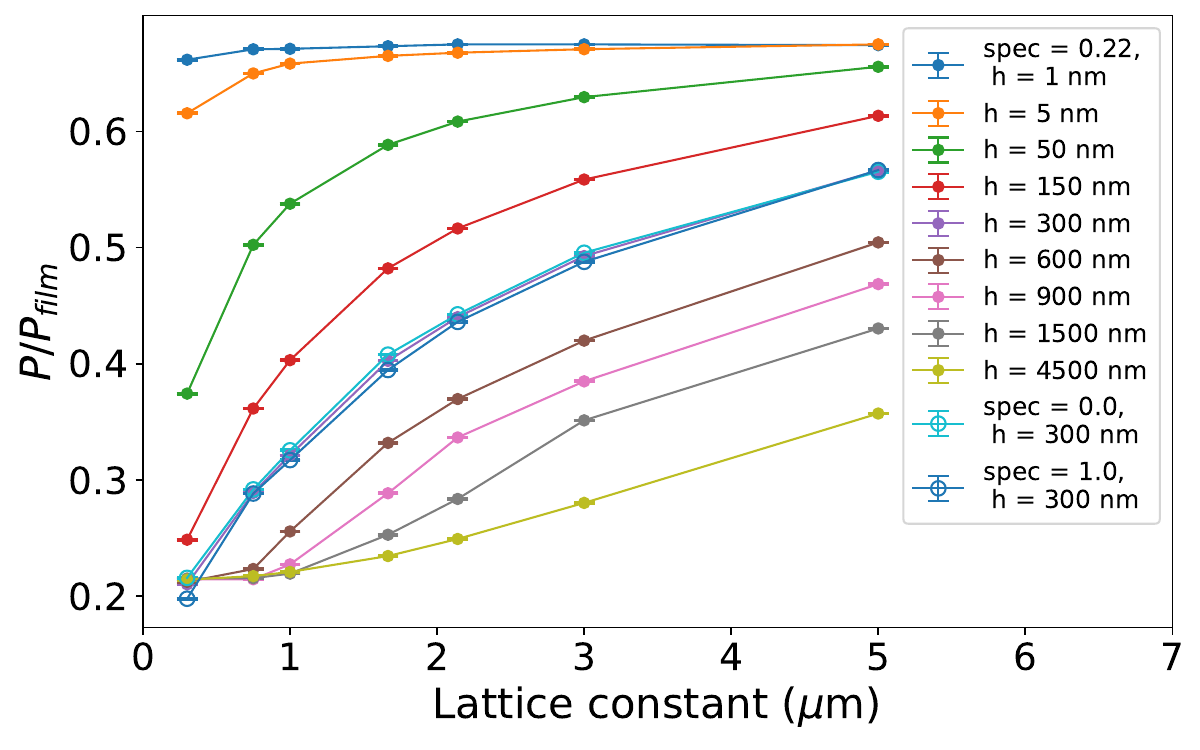}
    \caption{{\bf The effect of changing the the pillar height and the sidewall roughness on the Monte Carlo simulations.} All other parameters of the simulated PnC were the same as for the measured devices, with the sidewall specularity kept at $p = 0.22$ for the pillar heights $h \ne 300$ nm. The other specularity values were kept constant, with $p=1$ for the \SiN membrane, $p=0.99$ for the Al-\SiN interface, and $p=0.96$ for the top of the pillars.}
    \label{fig:monteplots}
\end{figure}

In this model, the sidewall roughness has a very minor effect on the thermal conductance of the PnC, even if the sidewall specularity is changed between its extremes of 0 and 1,  as shown in Fig. \ref{fig:monteplots}. 
 The pillar height $h$, on the other hand, has a significant effect on the incoherent transmission through the PnC (Fig. \ref{fig:monteplots}). At extremely low pillar heights $h \le 5$ nm, the conductance saturates to $\sim 67$ \% of the value for the membrane at larger lattice constants. Increasing the pillar height then steadily decreases the conductance, until at $h \geq 300$ nm the conductance again appears to settle at a minimum $P/P_{\textrm{film}} \approx 0.21$ at low periodicities, the lattice constant $a$ at which this saturation happens increasing with $h$. 

\subsection{Material density measurements with Rutherford Backscattering Spectrometry}
\label{Rutherford}

To accurately calculate the phonon dispersion relations of the PnCs with the FEM simulations, accurate values of the material parameters are required, without assuming values from the literature. One of them is the mass density $\rho$ of the materials in the PnC, the evaporated Al and LPCVD grown SiN$_\text{x}$, which were measured by Rutherford Backscattering Spectrometry (RBS). 

A test sample for 2 MeV $^4$He$^+$ RBS was prepared with an Al film evaporated under the same conditions as for the actual PnC samples. As the Al film thickness was accurately determined by atomic force microscopy (AFM), we could use the measured atomic areal density [1/m$^2$] to determine the mass density, giving $\rho_{\textrm{Al}} = 2.71 \pm 0.06 $ g/cm$^3$, which is for all intents and purposes the literature value for bulk Al density, 2.70 g/cm$^3$ \cite{CRC}. 

For silicon nitride, it is known that for low-stress LPCVD the mass density differs from bulk specimens \cite{Huszank_2016_SiN}. A \SiN film cut from the same wafer as the measured samples was also measured with RBS, yielding a density value of $\rho_{\textrm{SiN}} = 2.87 \pm 0.05$ g/cm$^3$. This is also in line with the results of Ref. \cite{Huszank_2016_SiN}, but clearly below the bulk value for stoichiometric Si$_3$N$_4$, 3.1 g/cm$^3$ \cite{Petersen,Cleland_kirja}. The element concentrations were 52.3 \% N and 47.7 \% Si, making the compound slightly Si rich compared to a stoichiometric Si$_3$N$_4$. For SiN$_\text{x}$, the film thickness was determined with spectroscopic ellipsometry.

\subsection{Elastic parameter determination via Brillouin Light Scattering}
\label{BLS}

The elastic properties of both the substrate and the pillars play a crucial role in the FEM analysis. Since we found that the mass density of the \SiN membrane differs from the bulk value, a similar deviation could also be expected for its elastic parameters. Furthermore, because the Al pillars were deposited by vacuum evaporation, it was necessary to account for the possibility that their elastic properties might differ from those reported in the literature for bulk Al.

The Young’s moduli and Poisson’s ratios of Al and \SiN were therefore determined using Brillouin light scattering (BLS). The measurements were based on the analysis of Lamb waves propagating in a pristine 320-nm thick \SiN membrane and subsequently in the same membrane coated with a 300-nm-thick Al layer. The experimentally obtained dispersion relations were fitted using FEM simulations, with the Young’s modulus and Poisson’s ratio treated as fitting parameters.

For Al, we obtained $E$ = 70 GPa and $\nu$ = 0.34, the bulk values for all intents and purposes. For SiN$_\text{x}$, the measured values were $E = 253.5$ GPa, and $\nu = 0.24$.

\subsection{Derivation of the equation for coherently emitted power}

In general, the heat flux $\dot{\bf Q}$ [W/m$^2$] carried by phonons is given by summing over all phonon modes $i$ the product of the energy$\hbar\omega_i$ and the group velocity $\partial\omega_i/\partial{\bf k}$  of each mode, weighed by the distribution function $n_i$, divided by the volume of the sample $V$:
\begin{equation*}
\dot{\bf Q} = \frac{1}{V}\sum_i\hbar\omega_in_i\frac{\partial\omega_i}{\partial{\bf k}}, 
\end{equation*}
where ${\bf k}$ is the wave vector. In our case, the different phonon modes are described by the 2D wave vector ${\bf k}$ and the branch index $j$, so that for a large PnC membrane with area $A$ and thickness $d_m$ (volume $V = Ad_m$), the summation over ${\bf k}$ can be converted into an integral:
\begin{equation*}
\sum_i \equiv \sum_j\frac{A}{(2\pi)^2}\int_K d{\bf k},
\end{equation*}
where the integral extends over the first Brillouin zone $K$ of the PnC. To calculate the emitted power from a narrow heater wire on top of the membrane, we assume that the emitted phonons are populated according to a local quasi-equilibrium, described by the Bose-Einstein distribution $n_B(\omega_j({\bf k}),T)$ with a heater temperature $T$. Another assumption is that each heater wire length element $dl$ emits freely in all directions without self-shadowing effects. This assumption is justified by the fact that the heater wire is narrow and is located only on the top surface of the membrane, instead of permeating the inside membrane volume under the heater. In practice, this assumption is taken into account by writing that the power emitted by a wire element $dl$ into each 2D mode $dP_i$ does not depend on the directional cosine with respect to the surface normal of the perpendicular plane of the membrane, $dA_n = d_m dl$, as was assumed before \cite{Zen_Comms_2014,Tian_maasilta_2019}. We can then write
$dP_i =  |\dot{\bf Q_i}|dA_n$, which, by integrating over the heater length and noting that the factors of area, volume and thickness cancel, $A d_m/V = 1$,  yields Eq. \eqref{eq:emitted_power_uus}:
\begin{equation*}
\begin{aligned}
 P &= \sum_i\int dP_i\\ 
 &= \sum_j\int dl \frac{1}{(2\pi)^2}\int_K d{\bf k}\hbar\omega_j({\bf k}) n_B(\omega,T)\left |\frac{\partial\omega_j({\bf k})}{\partial{\bf k}}\right |,  
\end{aligned}
\end{equation*}
as the integration over the heater length gives just the total length of the heater $\int dl = L$.



\section*{Code Availability}

The Monte Carlo code \cite{PhononRT} used for incoherent simulations is available on GitLab \\ (\url{https://gitlab.com/nanopnc/phonon_rt}). The coherent simulation code is available from the corresponding authors.

\bibliography{sources}

\begin{acknowledgments}
This work 
was supported by the Research Council of Finland Projects 
No. 341823 and No. 359240 (the Finnish Quantum Flagship, 
University of Jyväskylä).
The computational facilities provided 
by the CSC-IT Center for Science Ltd. are acknowledged. A.K. and B.G. acknowledge the National Science Centre of Poland (NCN) by the OPUS grant 2021/41/B/ST5/03038. 
\end{acknowledgments}

\section*{Author contributions}
I.M. and T.K. conceived the work, wrote the manuscript and interpreted the experimental findings. T.P. performed all the numerical FEM computations and created the Monte Carlo code. T.L. developed the heater-thermometer geometry and the fabrication process for it. Fabrication of the devices and the low-temperature experiments were performed by T.K. RBS measurements and analysis were performed by M. K. B.G. and A.K. performed the BLS measurements and data analysis.


\end{document}